\documentclass[sn-nature]{sn-jnl}

\usepackage{graphicx}%
\usepackage{multirow}%
\usepackage{amsmath,amssymb,amsfonts}%
\usepackage{amsthm}%
\usepackage{mathrsfs}%
\usepackage[title]{appendix}%
\usepackage{xcolor}%
\usepackage{textcomp}%
\usepackage{manyfoot}%
\usepackage{booktabs}%
\usepackage{algorithm}%
\usepackage{algorithmicx}%
\usepackage{algpseudocode}%
\usepackage{listings}%
\usepackage{caption}
\usepackage{array}
\usepackage{braket}

\theoremstyle{thmstyleone}%
\theoremstyle{thmstyletwo}%

\theoremstyle{thmstylethree}%

\begin{document}

\title[Article Title]{Direct observation of strain and confinement shaping the hole subbands of Ge quantum wells}


\author*[1,2]{\fnm{Enrico} \sur{Della Valle}}\email{denrico@ethz.ch}

\author[3]{\fnm{Arianna} \sur{Nigro}}

\author[4,5]{\fnm{Miki} \sur{Bonacci}}

\author[4,5]{\fnm{Nicola} \sur{Colonna}}
\author[3]{\fnm{Andrea} \sur{Hofmann}}

\author[4,5,6]{\fnm{Michael} \sur{Schüler}}

\author[4,5,7,8]{\fnm{Nicola} \sur{Marzari}}

\author[3,9]{\fnm{Ilaria} \sur{Zardo}}\email{ilaria.zardo@unibas.ch}

\author[2]{\fnm{Vladimir N.} \sur{Strocov}}\email{vladimir.strocov@psi.ch}

\affil*[1]{\orgdiv{Department of Physics and Quantum Center}, \orgname{Eidgenossische Technische Hochschule Zürich}, \orgaddress{\city{Zürich}, \postcode{8093}, \country{Switzerland}}}

\affil[2]{\orgdiv{Center for Photon Science}, \orgname{Paul Scherrer Institut}, \orgaddress{\city{Villigen}, \postcode{5232}, \country{Switzerland}}}

\affil[3]{\orgdiv{Department of Physics}, \orgname{University of Basel}, \orgaddress{\city{Basel}, \postcode{4056}, \country{Switzerland}}}

\affil[4]{\orgdiv{PSI Center for Scientific Computing, Theory and Data}, \postcode{5232} \orgaddress{\city{Villigen PSI}, \country{Switzerland}}}

\affil[5]{\orgdiv{National Centre for Computational Design and Discovery of Novel Materials (MARVEL)}, \orgname{École Polytechnique Fédérale de Lausanne}, \orgaddress{\city{Lausanne}, \postcode{1015}, \country{Switzerland}}}

\affil[6]{\orgdiv{Department of Physics}, \orgname{University of Fribourg}, \orgaddress{\city{Fribourg}, \postcode{1700}, \country{Switzerland}}}

\affil[7]{\orgdiv{Theory and Simulation of Materials (THEOS)}, \orgname{École Polytechnique Fédérale de Lausanne}, \orgaddress{\city{Lausanne}, \postcode{1015}, \country{Switzerland}}}

\affil[8]{\orgdiv{Theory of Condensed Matter, Cavendish Laboratory}, \orgname{University of Cambridge}, \orgaddress{\city{Cambridge}, \postcode{CB3 0US}, \country{United Kingdom}}}

\affil[9]{\orgdiv{Swiss Nanoscience Institute}, \orgname{University of Basel}, \orgaddress{\city{Basel}, \postcode{4056}, \country{Switzerland}}}


\abstract{
Germanium–silicon–germanium (Ge/Si$_x$Ge$_{1-x}$) heterostructures have emerged as a promising platform for hole-spin quantum technologies and high-mobility electronics, where strain and quantum confinement strongly reshape the Ge valence bands. However, the momentum-resolved valence-band structure of buried strained Ge quantum wells has so far been inferred only indirectly. Here we use soft X-ray angle-resolved photoemission spectroscopy (SX-ARPES) to directly probe the electronic structure of strained Ge quantum wells embedded in SiGe barriers. We resolve strain-split and size-quantized valence subbands, determine their heavy-hole, light-hole and split-off composition, and measure the valence-band offset at the Ge/SiGe heterojunction. Comparison with \textit{ab initio} calculations shows that an accurate description requires explicit inclusion of the confinement potential imposed by the SiGe barrier, which plays a decisive role in determining the dispersion, ordering and mixing of the hole states. Our results provide the first direct experimental picture of how strain and confinement determine the valence-band structure of Ge quantum wells, establishing a foundation for predictive modelling of hole-spin qubits and high-mobility devices based on group-IV heterostructures.
}

\maketitle

\section{Introduction}\label{sec1}
Ge has been the material of choice for building the first transistor, for high-performance amplifiers\
\cite{Xue2021}, and has recently become one of the most promising materials to build arrays of spin qubits \cite{hendrickxFastTwoqubitLogic2020, hendrickxFourqubitGermaniumQuantum2021a, hsiaoExcitonTransportGermanium2024, zhangUniversalControlFour2025}. Furthermore, Ge has become a viable candidate for combined superconductor-semiconductor devices \cite{lakicQuantumDotGermanium2025, tosatoHardSuperconductingGap2023, hendrickxBallisticSupercurrentDiscretization2019, aggarwalEnhancementProximityinducedSuperconductivity2021}, with the prospect of building superconducting transistors or various types of qubits \cite{sagiGateTunableTransmon2024b, kiyookaGatemonQubitGermanium2025a, laubscherGermaniumbasedHybridSemiconductorsuperconductor2024, spethmannHighfidelityTwoqubitGates2024, luethiMajoranaBoundStates2023}. These advancements stem from the favorable properties of the holes confined in Ge quantum wells \cite{Scappucci2020}, which include a low effective mass \cite{Lodari2019}, the possibility for fast, electric qubit manipulation \cite{hendrickxFastTwoqubitLogic2020, jirovecSinglettripletHoleSpin2021, Sarkar2023, Stano2025, Miller2022, Drichko2018}, high mobility \cite{Myronov2023}, the Fermi level pinning close to the valence band and the absence of piezoelectricity, valleys and contact hyperfine interaction \cite{Burkard2023}. As extracted from density functional theory (DFT) calculations \cite{Terrazos2021} and indirectly inferred from transport experiments \cite{Jirovec2021, Mizokuchi2017, sammak2019}, the band structure of the holes in 2D Ge/SiGe heterostructures yields heavy-hole-like (HH-like) ground states, and the light-hole (LH) and the spin-orbit split-off (SO) bands have a higher hole energy. While the field has been relying on these assumptions, direct experimental measurements of the band structure and the HH/LH/SO character of the subbands have so far been missing.

\begin{figure}[H]
\centering
\includegraphics[width=1.0\textwidth]{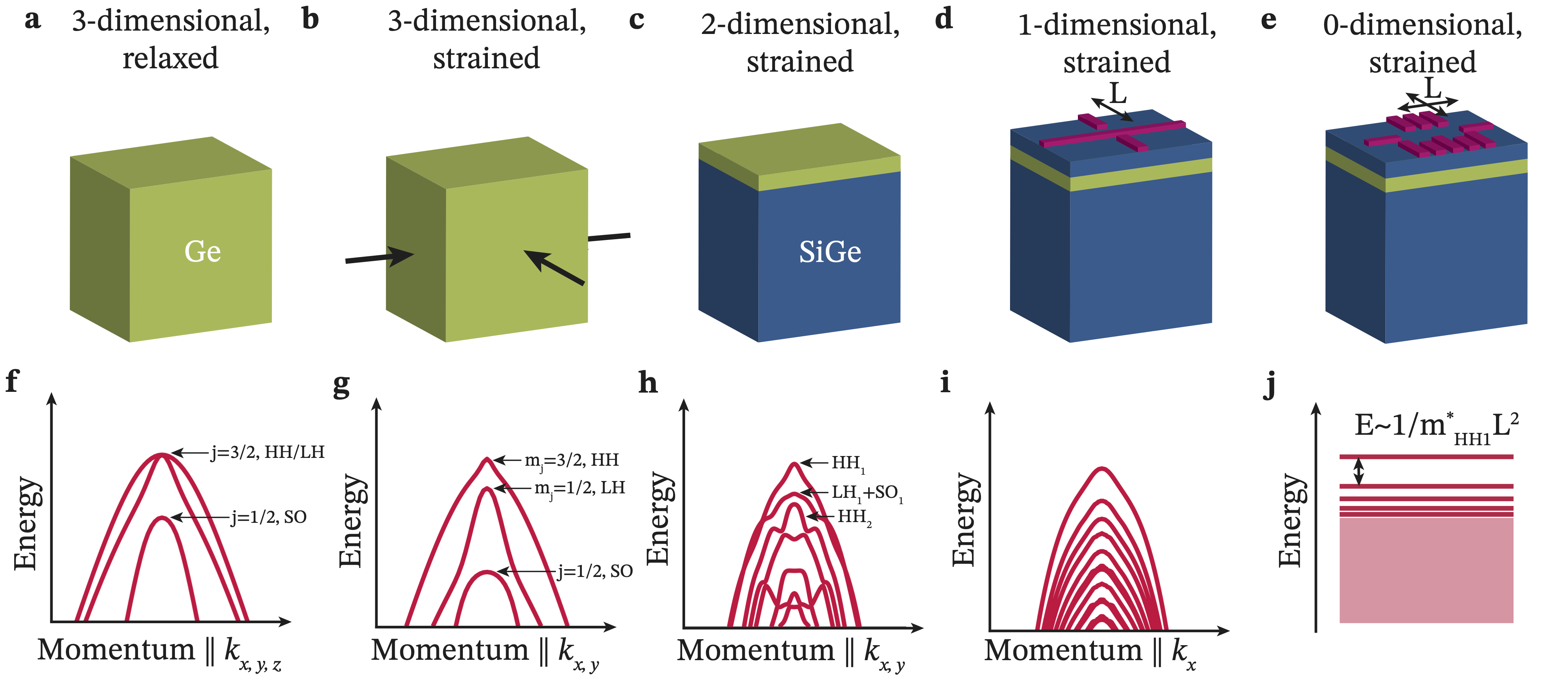}
\caption{\textbf{Strain and size quantization effects on energy levels.}
Schematics illustrating how strain and size quantization affect the momentum-resolved band structure of germanium. (a) Bulk (3D) germanium and its corresponding band structure in (f). (b) Biaxial strain in bulk Ge (3D strained) lifts the degeneracy between HH and LH bands at $\Gamma$, as shown in (g). (c) Quantization along one axis (quasi-2D strained) leads to the formation of subbands with mixed HH/LH/SO character, as shown in (h). (d) Further confinement in an additional direction gives rise to the schematic energy-level structure shown in (i). Finally, confinement in all directions (0D strained) results in quantum dots, characterized by the underlying energy levels in (h).}
\label{fig1}
\end{figure}

Figure~\ref{fig1} summarizes the effects of strain and confinement onto the Ge band structure. In relaxed bulk Ge, the valence states are derived from $p$-like HH, LH, and SO bands with a substantial spin–orbit splitting of 260~meV~\cite{Scappucci2020} (Figs.~\ref{fig1}(a) and (f)). Compressive biaxial strain lifts the HH–LH degeneracy and reduces the effective mass of the topmost HH-like band by nearly a factor of five~\cite{Terrazos2021}, as shown in Figs.~\ref{fig1}(b) and (g). Yet, once translational symmetry is broken along the growth direction, as is the case for quantum wells, the bulk HH, LH, and SO labels become only approximate: mixing between these states, mediated by confinement and interface potentials, modifies their band character, dispersion, and spin properties, as schematically shown in Figs.~\ref{fig1}(c) and (h). Because the energy-level spacing in quantum wires and quantum dots scales inversely with the effective mass and the square of the lateral confinement~\cite{Winkler2003}, this strain- and confinement-induced band reshaping directly influences the design and performance of both one-dimensional gate-defined quantum wires~\cite{Mizokuchi2018} (Figs.~\ref{fig1}(d) and (i)) and zero-dimensional hole-spin qubits~\cite{Lodari2019} (Figs.~\ref{fig1}(e) and (j)).


In this work, we combine soft X-ray angle-resolved photoemission spectroscopy (SX-ARPES) with \textit{ab initio} DFT and tight-binding (TB) calculations to obtain a complete, momentum-resolved picture of the electronic structure of relaxed Ge and Ge/SiGe quantum wells. Our measurements directly resolve the strain- and size-induced modifications of the valence band, including the formation of subbands in the Ge QW, their mixed band character, and the valence-band offset at the Ge/SiGe interface with momentum resolution. By comparing experiment and theory, we show that a quantitative description of Ge QWs requires explicit inclusion of the SiGe barrier and its symmetry, strain, and confinement effects. These insights establish an experimentally validated framework for predictive band-structure engineering in Ge-based high-mobility devices and hole-spin qubit architectures.

\section{Results}\label{sec2}
\subsection{Strain control and characterization}\label{sec2.1}

\begin{figure}[H]
\centering
\includegraphics[width=1.0\textwidth]{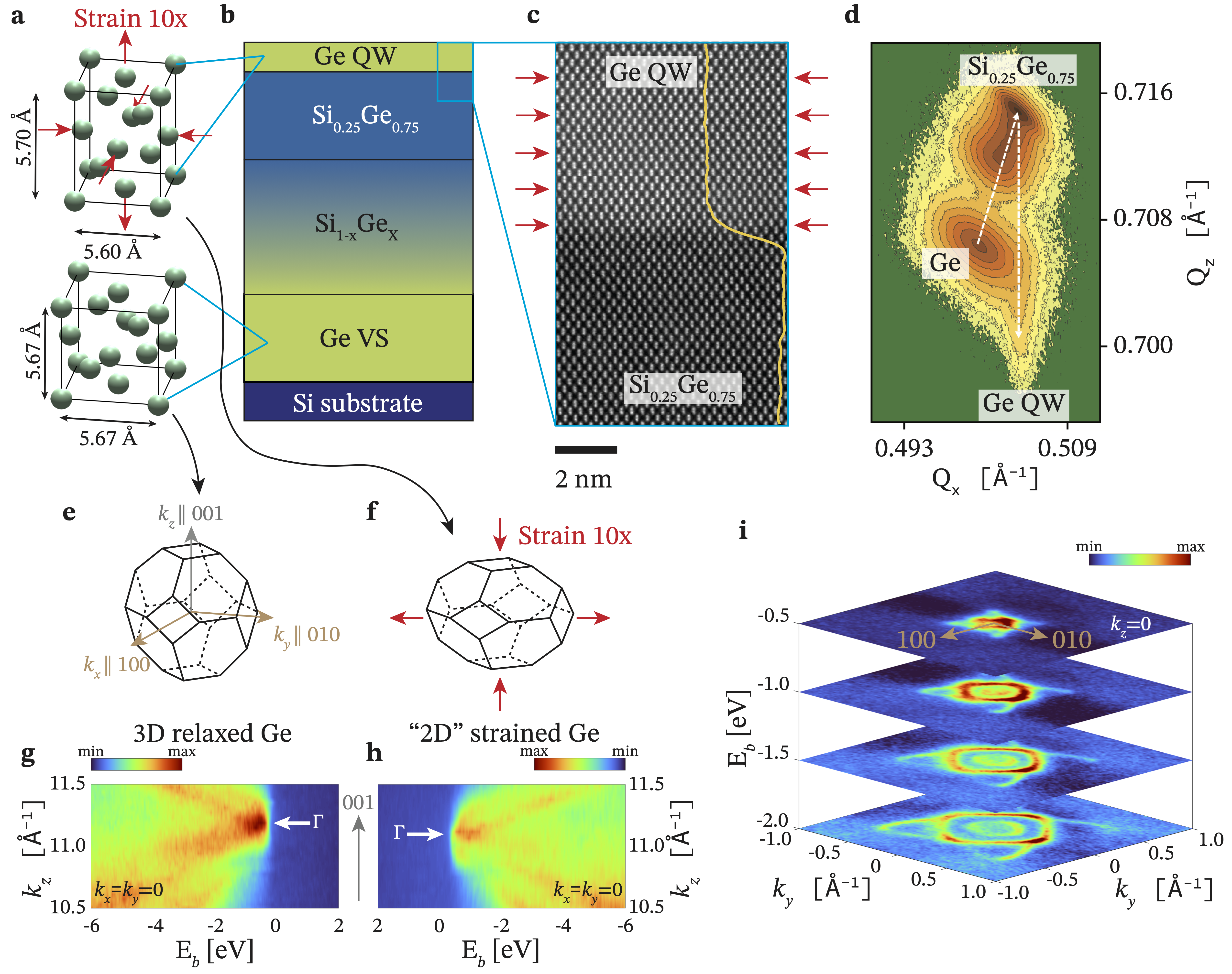}
\caption{\textbf{Heterostructure design and characterization.} (a) Unit cell of Ge under $\varepsilon = -10\%$ biaxial compressive strain (top), shown to illustrate the overall effect of strain, and relaxed Ge (bottom). (b) Schematic representation of the planar Ge/SiGe heterostructure. (c) Atomic-resolution high-angle annular dark-field (HAADF) scanning transmission electron microscopy image acquired at the interface between the Si\textsubscript{0.25}Ge\textsubscript{0.75} barrier and the Ge QW. The yellow curve corresponds to the HAADF line-profile intensity. (d) X-ray diffraction reciprocal space map of the (224) reflection of the heterostructure. (e) Brillouin zone of Ge showing the relevant in-plane ($k_x$, $k_y$) and out-of-plane ($k_z$) directions. (f) Brillouin zone under $\varepsilon = -10\%$ biaxial compressive strain, highlighting the effect of strain in reciprocal space. (g,h) Out-of-plane dispersion (at $k_x=k_y=0$) of relaxed bulk Ge and the strained Ge QW, respectively, as measured by SX-ARPES. (i) In-plane dispersion of the strained Ge QW as measured by SX-ARPES.}
\label{fig2}
\end{figure}

Planar Ge/SiGe heterostructures were grown via chemical vapor deposition (CVD) according to the design reported in Fig. \ref{fig2}(b). By following the reverse grading approach \cite{shah2008}, a relaxed Ge film and a reverse linearly graded Si\textsubscript{1-x}Ge\textsubscript{x} alloy were deposited on a Si(100) substrate, followed by a Si\textsubscript{0.25}Ge\textsubscript{0.75} barrier and a 5 nm thick Ge QW (see Methods).

The crystalline quality of the different layers was assessed via high-angle anular dark-field (HAADF) scanning transmission electron microscopy (STEM). Fig. \ref{fig2}(c) shows the atomically resolved HAADF image taken at the interface between the Si\textsubscript{0.25}Ge\textsubscript{0.75} barrier and the Ge QW. Both materials are single crystalline and free of defects, as a result of the successful implementation of the reverse grading approach, with a transition region between the two layers limited to a few monolayers. A direct quantification of the sharpness at the interface, performed by fitting the normalized intensity profile, corresponding to the yellow curve in Fig. \ref{fig2}(c), with an error function \cite{clark2008}, reveals an upper bound as low as 0.4~nm, significantly lower than what is currently reported in literature \cite{bashir2018,ciano2019,grange2020,zhang2022}.

Strain and composition were determined from asymmetric (224) X-ray diffraction reciprocal-space mapping (Fig.~\ref{fig2}(d)). Three dominant peaks are observed, corresponding to the relaxed Ge virtual substrate, the Si\textsubscript{0.25}Ge\textsubscript{0.75} barrier, and the Ge QW. The diffuse scattering between the Ge and Si\textsubscript{0.25}Ge\textsubscript{0.75} peaks originates from the graded Si\textsubscript{1–x}Ge\textsubscript{x} buffer. The Si\textsubscript{0.25}Ge\textsubscript{0.75} peak position indicates a Ge content of 76\% with a residual in-plane tensile strain of only 0.08\%. The vertical alignment of the Ge QW and Si\textsubscript{0.25}Ge\textsubscript{0.75} peaks demonstrates pseudomorphic growth of the QW, corresponding to a compressive strain of $-0.81\%$, in agreement with previous reports~\cite{sammak2019,stehouwer2023}. This precise strain control is essential for achieving the targeted modifications of the valence-band structure explored in the following sections. 

Because angle-resolved photoemission spectroscopy (ARPES) probes only the top few nanometers of a sample (even in the soft X-ray regime) it is essential to verify that the near-surface layers of our heterostructure remain strained. Soft X-ray ARPES (SX-ARPES) provides a direct way to assess this, as it gives access to the momentum-resolved electronic structure along all three dimensions in momentum space \cite{Strocov2012,Lev2015}.

Fig.~\ref{fig2}(e) illustrates the Brillouin zone of Ge, with the in-plane momentum components $k_x$ and $k_y$ indicated in brown and the out-of-plane component $k_z$ in gray. Under in-plane biaxial compression (Fig.~\ref{fig2}(a), top), the real-space lattice expands along the growth direction to approximately conserve volume, effectively compressing the Brillouin zone along $k_z$. SX-ARPES, with its increased probing depth at $h\nu \sim 500-1000$~eV, also provides improved $k_z$ resolution due to the reduced intrinsic broadening imposed by the Heisenberg uncertainty principle \cite{Strocov2003}. Consequently, by measuring the dispersion at $k_x = k_y = 0$ while varying the photon energy, we can precisely identify the $k_z$ values at which the valence-band maximum (which corresponds to the $\Gamma$ point) is reached and thereby determine the out-of-plane lattice constant of the probed layers.

Figs.~\ref{fig2}(g)–(h) show that the photon energy required to access the $\Gamma$ point differs for relaxed and strained Ge: 461~eV for the bulk-relaxed sample and 454~eV for the strained quantum well. Using the free-electron final-state approximation at normal emission ($k_x = k_y = 0$) and at the valence-band maximum ($E_b \approx 0$), we write $k_z = \frac{\sqrt{2 m_0}}{\hbar}\sqrt{h\nu + V_{000}} + k^{ph}_{\perp},$ where $k^{ph}_{\perp}$ is the photon momentum projected along the surface normal. From this analysis, we extract an inner potential of $V_{000} = 9$~eV, which reproduces the expected periodicity $k_z = n \frac{2\pi}{c}$ and is consistent with literature values.
Applying the same procedure to the strained case yields $c = 5.70$~\AA{}, fully consistent with expectations and with the X-ray diffraction data discussed above. This confirms that the near-surface layers probed by SX-ARPES remain pseudomorphically strained.

Finally, the in-plane constant-energy map in Fig.~\ref{fig2}(i) exhibits sharp dispersive features over a wide range of binding energies, indicating excellent crystalline order at the surface, an essential requirement for reliable momentum-resolved spectroscopy.

\subsection{Ge/SiGe interface}\label{sec2.2}

\begin{figure}[H]
\centering
\includegraphics[width=1\textwidth]{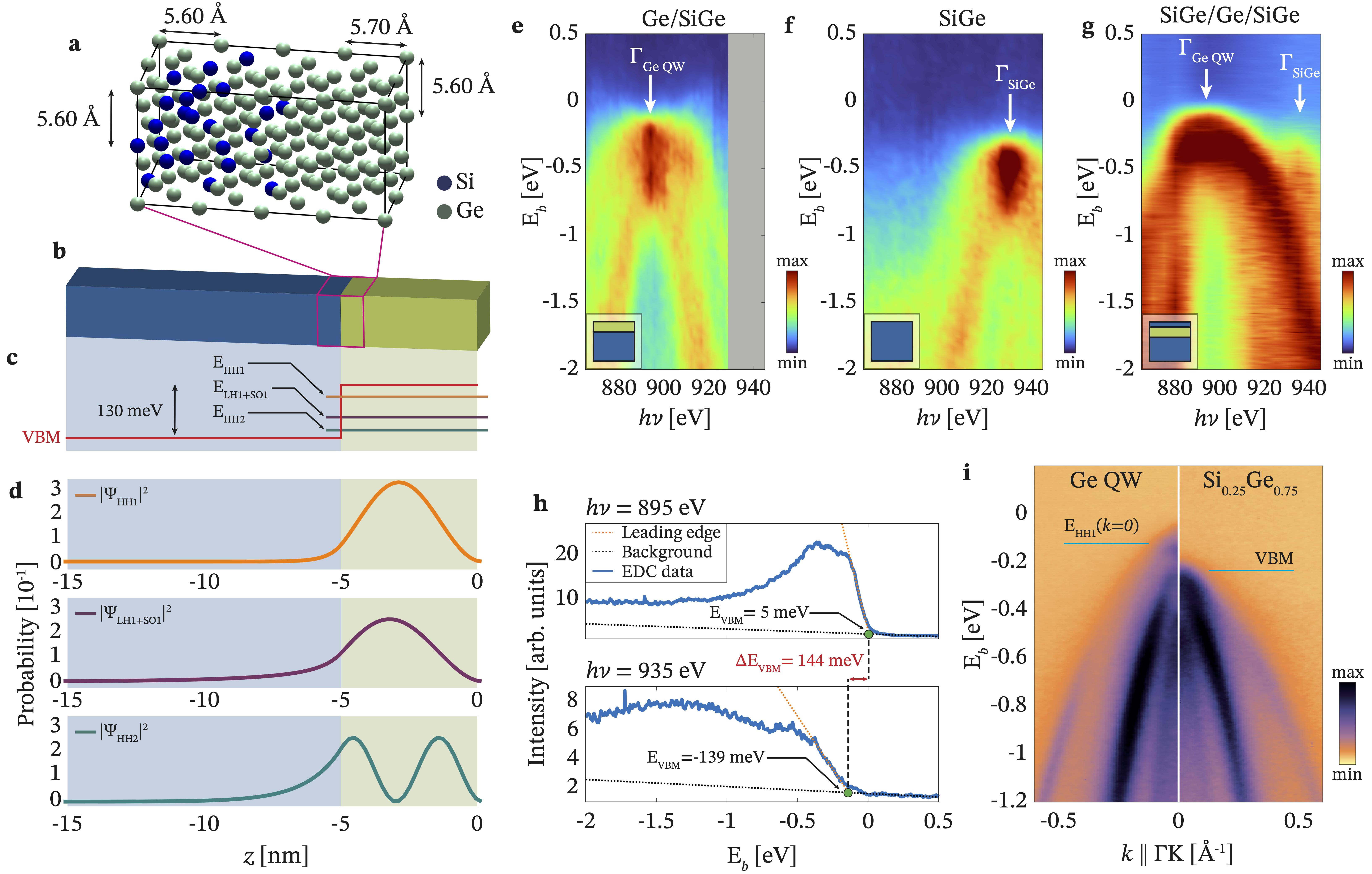}
\caption{\textbf{Valence-band offset at the Ge/SiGe interface.}
(a) Atomic structure of the Si\textsubscript{0.25}Ge\textsubscript{0.75}/Ge interface, with the relevant in-plane and out-of-plane lattice constants indicated. (b) Schematic of the 1D supercell used to calculate the quantum-well states in the Ge QW using a self-consistent Poisson–Schrödinger algorithm~\cite{Birner2007}. (c) Corresponding valence-band profile across the heterojunction, together with the energies of the bound states. (d) Real-space probability distributions of the three uppermost bound states in the Ge QW. (e,f) $k_z$-dependent photoemission intensity at $k_x = k_y = 0$ for the Ge QW and bulk Si\textsubscript{0.25}Ge\textsubscript{0.75}, respectively, with the corresponding valence-band maxima emphasized by white arrows. (g) $k_z$-dependent ARPES map at $k_x = k_y = 0$ for a Si\textsubscript{0.25}Ge\textsubscript{0.75}/Ge/Si\textsubscript{0.25}Ge\textsubscript{0.75} heterostructure, simultaneously showing the valence-band maxima of the Ge QW and Si\textsubscript{0.25}Ge\textsubscript{0.75} at their respective $k_z$ positions, as indicated by the two white arrows. (h) EDCs extracted from (g) at the $h\nu$ values 895 eV and 935 eV. The leading band edge method is used to extract the VBM position of both materials, thereby obtaining the band offset at the interface. (i) Aligned Ge QW and Si\textsubscript{0.25}Ge\textsubscript{0.75} band structures, including dispersion along the $\Gamma$K direction.}
\label{fig3}
\end{figure}

The interface between Ge and SiGe is fundamental to all device architectures that rely on the accumulation of a two-dimensional hole gas (2DHG) in Ge/SiGe heterostructures. A key requirement for hole confinement is the valence-band offset between the two materials: the valence-band maximum of Ge lies higher in energy than that of SiGe, as illustrated in Fig.~\ref{fig3}(c). The magnitude of this offset depends on the Si content of the barrier layer. Here, we directly probe the valence-band offset between Si$_{0.25}$Ge$_{0.75}$ and Ge using SX-ARPES, which provides the necessary probing depth to access both materials simultaneously.

Our approach builds on the strategy used in Fig.~\ref{fig2} to determine the strain state near the surface. Because strained Ge and Si$_{0.25}$Ge$_{0.75}$ exhibit different out-of-plane lattice constants (Fig.~\ref{fig3}(a)), their Brillouin zones have different size along $k_z$. We therefore begin by independently measuring the $k_z$-dependent band dispersions of the strained Ge QW and of Si$_{0.25}$Ge$_{0.75}$ on separate reference samples, shown in Fig.~\ref{fig3}(e) and (f), respectively.

To measure the band offset directly, we fabricated a SiGe/Ge/SiGe heterostructure containing a thin ($\sim 1.5$~nm) SiGe overlayer above the Ge QW. At photon energies around $h\nu \sim 900$~eV, the probing depth of SX-ARPES is sufficient to detect photoemission from both the SiGe cap and the underlying Ge QW. The resulting $k_z$-dependent spectra at $k_x = k_y = 0$, shown in Fig.~\ref{fig3}(g), simultaneously reveal the VBM of strained Ge at lower $k_z$ (i.e., lower $h\nu$) and the VBM of Si$_{0.25}$Ge$_{0.75}$ at higher $k_z$ and lower binding energy. From the energy separation between these two features, we extract a valence-band offset of $(144 \pm 30)$~meV using the leading-edge method~\cite{Constantinou2025}, as illustrated in Fig.~\ref{fig3}(h). As shown in Fig.~\ref{fig3}(g), the signal originating from the ultrathin SiGe overlayer is less intense than that from the buried Ge layer. This reduction is most likely due to static disorder in the few-monolayer-thick SiGe cap near the interface, which is known to suppress dispersive features in ARPES~\cite{DellaValle2026}.

Fig.~\ref{fig3}(h) compares the band structures of the Ge QW (left, measured on the sample shown in Fig.~\ref{fig3}(e)) and Si$_{0.25}$Ge$_{0.75}$ (right, measured on the sample shown in Fig.~\ref{fig3}(f)), aligned according to the experimentally determined offset. To support the interpretation of the ARPES spectra, we simulate the heterostructure using a 1D Poisson-Schrödinger solver~\cite{Birner2007} and a tight-binding (TB) calculation derived from \textit{ab initio} simulations, as discussed in more depth in the next subsection. As introduced in Figs.~\ref{fig3}(c) and (d), the system is modeled using a 1D slab containing a 5~nm Ge QW and a 10~nm Si$_{0.25}$Ge$_{0.75}$ barrier. For the TB calculation, the barrier is approximated as strained Ge with an applied valence-band shift of 170~meV relative to the Ge QW. Both the 1D Poisson-Schrödinger and TB calculations yield three bound QW states, whose squared envelope wave functions are plotted against $z$ in Fig.~\ref{fig3}(d). In the next subsection, we show that explicitly including the SiGe barrier in the theoretical modeling of the Ge QW is essential for correctly interpreting its measured $\mathbf{k}$-resolved band structure.

\subsection{Strain and confinement effects on the Ge band structure}\label{sec2.3}

For both strained and relaxed Ge(001), varying the photon energy between 350~eV and 1000~eV allows access to the $\Gamma$~point in three successive Brillouin zones. For strained Ge at $k_x = k_y = 0$, these occur at $h\nu = 454$~eV, 651~eV, and 890~eV, respectively. To optimize both energy and momentum resolution, we performed all measurements at $h\nu = 454$~eV. At this photon energy, the out-of-plane momentum resolution, estimated from the inelastic mean free path~\cite{Tanuma1994}, is $\Delta k_z = 0.08~\text{\AA}^{-1}$. The overall energy resolution, obtained by fitting the Fermi edge of Au measured under identical conditions, is $\Delta E = 47$~meV, and the in-plane momentum resolution is $\Delta k_x = 0.02~\text{\AA}^{-1}$.

\begin{figure}[H]
\centering
\includegraphics[width=1.0\textwidth]{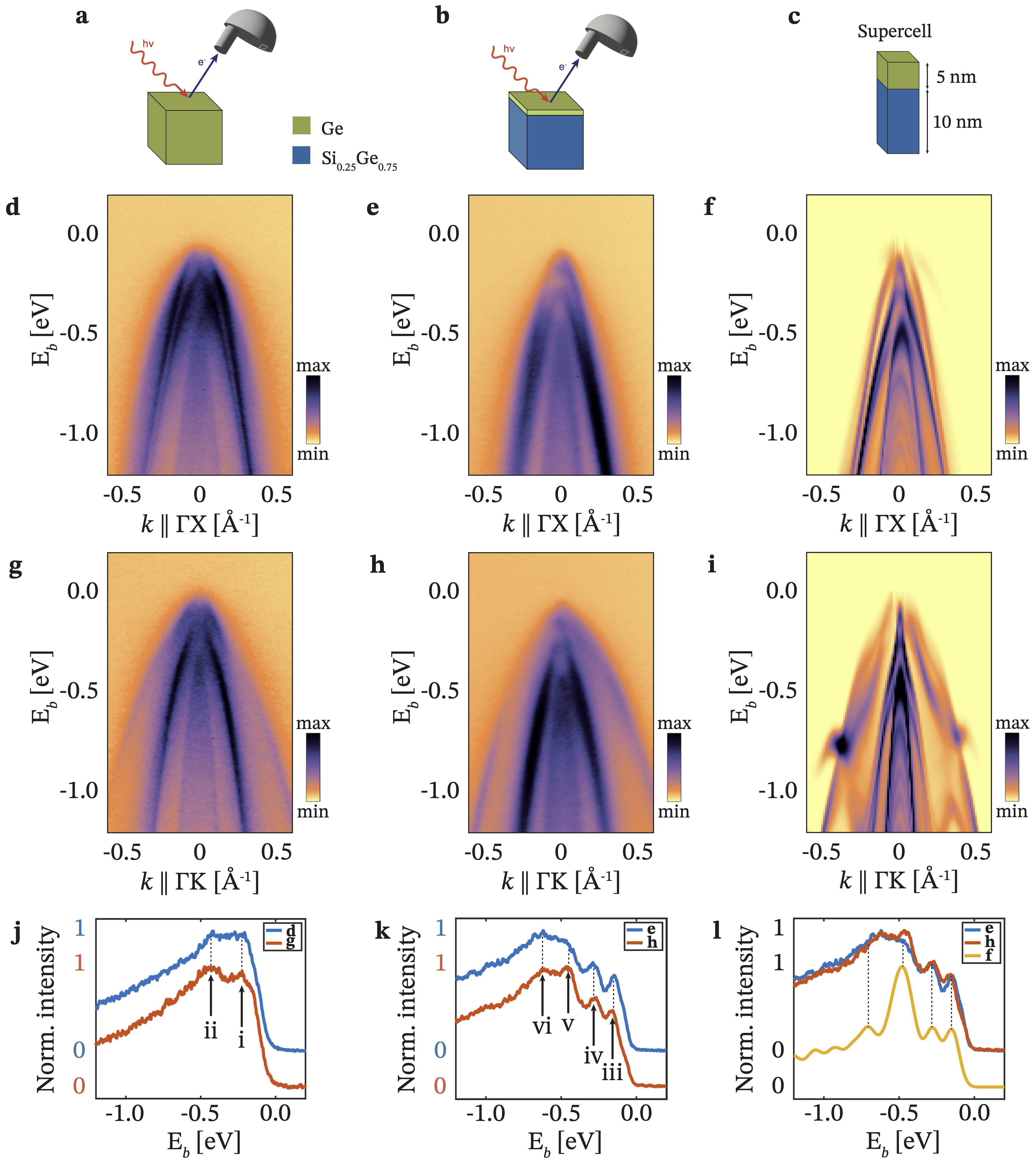}
\caption{\textbf{Electronic structure of 3D and quasi-2D strained Ge.} (a,b) Schematics of the experimental setup, including the sample structure. (c) Supercell used in the \textit{ab initio} tight-binding calculations. (d,g) Electronic band structure of bulk (3D) Ge measured by SX-ARPES at $h\nu=460$ eV along the $\Gamma$X and $\Gamma$K high-symmetry directions, respectively. (e,h) Electronic band structure of quasi-2D compressively strained Ge measured by SX-ARPES at $h\nu=454$ eV along the $\Gamma$X and $\Gamma$K high-symmetry directions, respectively. (f,i) \textit{Ab initio} tight-binding calculations corresponding to the spectra shown in (e) and (h), using the supercell in (c). (j,k) EDCs at the $\Gamma$ point for bulk Ge and quasi-2D compressively strained Ge, with all relevant states highlighted by arrows. (l) Comparison between the experimental EDCs of quasi-2D compressively strained Ge and the corresponding tight-binding calculation.}
\label{fig4}
\end{figure}

To measure along the $\Gamma$X ($\Gamma$K) direction, the sample was oriented such that the $[100]$ ($[110]$) crystallographic direction lay within the measurement plane. The background-subtracted spectra are shown in Fig.~\ref{fig4}: the left column displays the data from bulk relaxed Ge, while the middle column shows the strained Ge QW. The bottom panels of Fig.~\ref{fig4} present energy distribution curves (EDCs) extracted at the $\Gamma$~point using a momentum-integration window of $\pm 0.01~\text{\AA}^{-1}$.

The EDCs for bulk relaxed Ge (Fig.~\ref{fig4}(j)) exhibit two distinct peaks. The peak closest to the Fermi energy originates from the degenerate HH and LH bands, which remain four-fold degenerate with total angular momentum $j = 3/2$~\cite{Terrazos2021}. The lower-energy peak corresponds to the SO band. In contrast, the strained Ge QW displays four prominent peaks. This behavior deviates both from the bulk relaxed case and from the simple bulk strained-Ge scenario shown in Fig.~\ref{fig1}(g), where strain only lifts the HH–LH degeneracy.

Computational investigation of the electronic structure of 3D strained Ge was performed by means first-principles calculations. Specifically, DFT was employed, taking into account spin-orbit coupling effects. The strained structure was constructed starting from the one provided by Terrazos \textit{et al.}~\cite{Terrazos2021}. The resulting valence band structure, reported in Fig.~\ref{figS2}, yields HH–LH and HH–SO splittings of 159~meV and 608~meV, respectively, in good agreement with previous reports~\cite{Terrazos2021}. Details of the computational setup are provided in Methods. 
These bulk calculations alone cannot account for the four-band structure observed experimentally, indicating that additional quantization effects must be considered.

Size quantization arises because the valence-band offset discussed in the previous subsection produces an energy barrier on the SiGe side, confining states within $\sim 144$~meV of the VBM to the Ge QW. In standard SiGe/Ge/SiGe wells, confinement occurs on both sides; in our Ge/SiGe structure, the vacuum termination introduces an even more restrictive confinement barrier at the surface. To quantify this situation, we compute the ARPES spectra using the supercell geometry shown in Fig.~\ref{fig4}(c). For computational efficiency, instead of explicitly treating Si$_{0.25}$Ge$_{0.75}$, the barrier is approximated as strained Ge ($\varepsilon = -1$\%) with a 170~meV valence-band shift.

The ARPES spectra are calculated using a Wannier-ARPES framework: the hybrid-functional DFT bands and wave functions mentioned above are Wannierized, yielding a tight-binding Hamiltonian that is stacked to form the slab. Diagonalizing this Hamiltonian provides the eigenstates of the supercell, and the ARPES intensity is obtained via Fermi's golden rule using evanescent plane-wave final states and a dipole operator appropriate for circular polarization. The resulting spectra are shown in the right column of Fig.~\ref{fig4}.

Crucially, the \textit{ab initio} calculations reproduce the four-peak structure observed experimentally in the EDCs (Fig.~\ref{fig4}(k)). In the next subsection, we build on this analysis by examining the HH/LH/SO character of these quantized states.

\subsection{Valence-band mixing}\label{sec2.4}
Another important consequence of breaking translational symmetry along the growth direction is the hybridization between HH, LH, and SO states. This mixing is not only of fundamental interest, but also has direct implications for device technologies that rely on holes in strained Ge QWs. For example, high-hole-mobility transistors may exhibit an effective mass that varies with carrier concentration, partly because the character of the occupied subband evolves with momentum.

Similarly, additional confinement in the in-plane directions to define spin qubits, as illustrated in Fig.~\ref{fig1}(e), is sensitive to valence-band mixing for two reasons. First, depending on the physical size of the qubit, the dominant band character inherited from the quasi-2D valence-band structure, and thus the effective mass, may vary, modifying the energy-level spacing. Second, the spin properties of hole states can also be altered by their band composition.

\begin{figure}[H]
\centering
\includegraphics[width=1.0\textwidth]{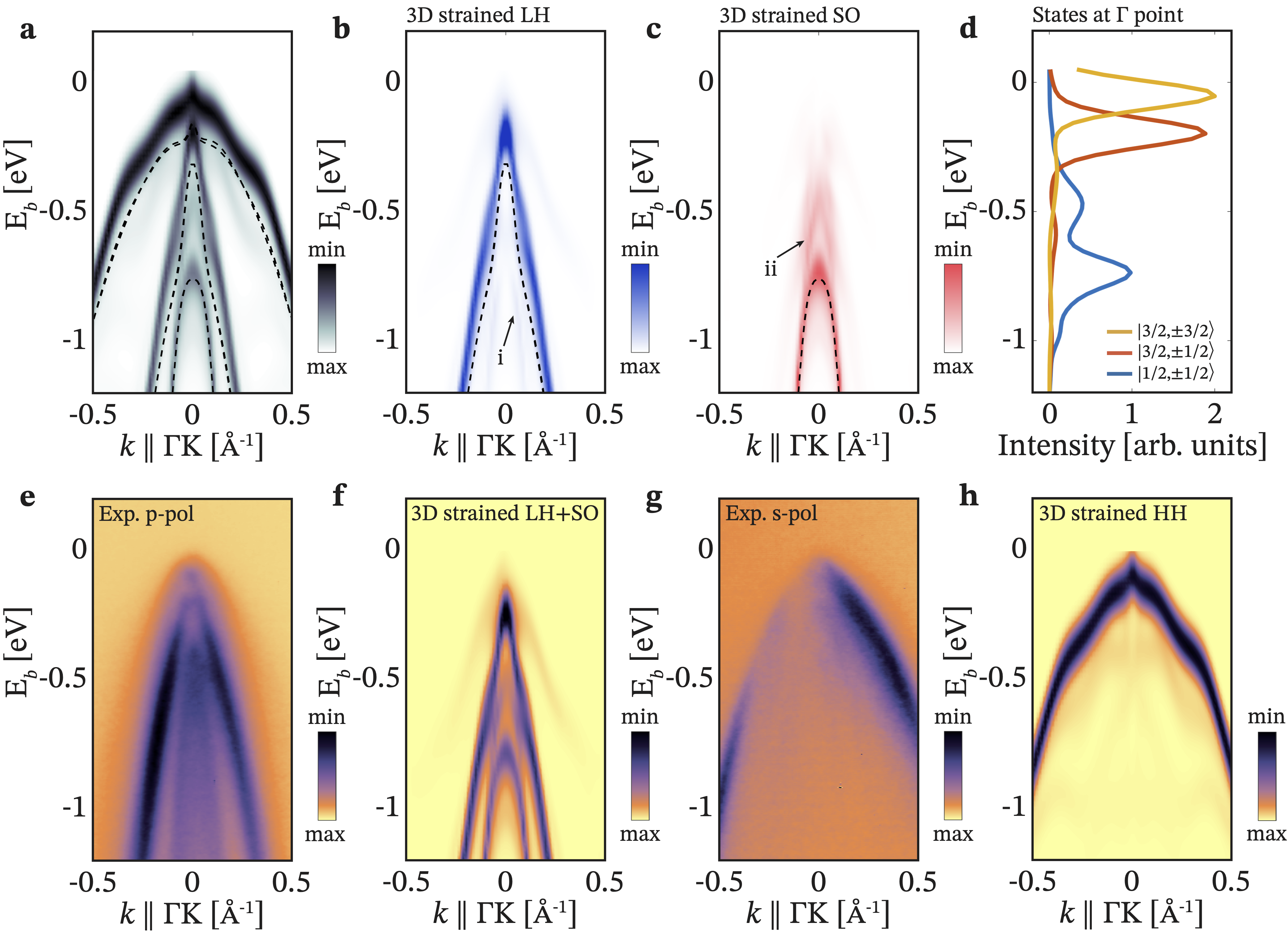}
\caption{\textbf{Strain- and size-quantization-induced HH/LH/SO mixing.}
(a) Spectral function along the $\Gamma$K high-symmetry direction, calculated for the supercell shown in Fig.~\ref{fig3}(b) and projected onto all bulk bands of strained Ge within the top 5~nm of the Ge quantum well. Dashed lines indicate the corresponding HSE bulk band structure of strained Ge. Spectral functions projected onto the (b) LH, (c) SO, and (h) HH bulk bands of strained Ge, respectively; dashed lines again show the HSE bulk bands. (d) Intensity at $k=0$, extracted from panels (b), (c), and (h), revealing the band character and ordering of the subbands. (e) Experimental band structure measured by SX-ARPES along $\Gamma$K for the 5~nm Ge quantum well using linear vertical ($p$-polarized) light, and (f) the summed spectral function from panels (b) and (c). (g) Experimental band structure measured by SX-ARPES along $\Gamma$K using linear horizontal ($s$-polarized) light, and (h) the corresponding spectral function projected onto the HH bulk band of strained Ge.}
\label{fig5}
\end{figure}

Here, we use the supercell calculations introduced above to determine the valence-band character of the subbands, and SX-ARPES with linear polarization along the high-symmetry $\Gamma$K direction to experimentally verify these predictions. Figure~\ref{fig5}(a) shows the spectral function of the Ge QW projected onto the bulk strained bands, together with the corresponding HSE band structure of bulk strained Ge (dashed lines). A notable feature is that the calculated spectra of the QW are stretched along the energy axis relative to the bulk bands. This energy rescaling arises from the valence-band offset between Ge and the Si$_{0.25}$Ge$_{0.75}$ barrier, as well as from hybridization between their states, consistent with Fig.~\ref{fig3}(d).

In Figs.~\ref{fig5}(b,c,h), we show the spectral function of the top 5~nm of the Ge QW projected onto the bulk strained LH, SO, and HH bands, respectively. Although the strained LH, SO, and HH states do not strictly retain the same band character as their relaxed counterparts, at the $\Gamma$ point we can still, to a good approximation, identify HH as $\ket{3/2,\pm3/2}$, LH as $\ket{3/2,\pm1/2}$, and SO as $\ket{1/2,\pm1/2}$ \cite{Terrazos2021}.

From Fig.~\ref{fig5}(d), we see that the first occupied hole states are still $\ket{3/2,\pm3/2}$ (HH-like), as already inferred from indirect transport measurements. The second state is $\ket{3/2,\pm1/2}$, namely LH-like, while the remaining two states are $\ket{1/2,\pm1/2}$ and therefore SO-like. This allows us to assign the four states observed in Fig.~\ref{fig4}(k,l) at $k=0$. As soon as $k\neq0$, however, HH, LH, and SO mix. As before, the bulk strained band structure is overlaid as dashed lines. A key observation is that the HH-, LH-, and SO-projected spectral functions do not simply follow the bulk bands and are not confined to a single subband. Instead, they give rise to multiple dispersive features (see features~i and~ii in Fig.~\ref{fig5}). In Fig.~\ref{fig5}(c), the lowest subband closely follows the strained bulk SO band, supporting the assignment of the fourth peak in Fig.~\ref{fig4}(k) (feature~vi) to an SO-derived ``bulk'' subband.

Next, we exploit the fact that the bulk relaxed and bulk strained HH, LH, and SO bands transform according to the same irreducible representations along $\Gamma$K, although not at $\Gamma$. The HH band has even mirror parity under $[1\bar{1}0]$, whereas the LH and SO bands have odd mirror parity. As a result, ARPES measurements using $p$- or $s$-polarized light can selectively probe the corresponding HH and LH/SO components away from the $\Gamma$ point, allowing direct comparison with the HH-, LH-, and SO-projected spectral functions from the supercell calculations.

Using $p$-polarized light along $\Gamma$K (Fig.~\ref{fig5}(e)), we predominantly probe the LH+SO character, which we compare with the summed LH- and SO-projected spectral functions shown in Fig.~\ref{fig5}(f). The agreement is good: the SO-derived subband is clearly visible in both theory and experiment, and an additional subband appears in both datasets, although with different relative intensities. At the valence-band maximum, that is, at the $\Gamma$ point, the experimental spectrum shows a clear gap that is not present in the calculated spectral function. This discrepancy may arise because the selection rules exactly at $\Gamma$ differ from those along $\Gamma$K.

The SX-ARPES spectrum measured with $s$-polarized light is shown in Fig.~\ref{fig5}(g). Here, we observe not only the expected HH band at the top but also additional intensity at lower binding energies. Compared to the HH-projected spectral function, the cusp predicted at the $\Gamma$ point is not clearly resolved in the experimental data. This difference is consistent with the fact that the calculated spectral function contains all states localized within the Ge QW, whereas ARPES preferentially enhances states closest to the surface. As seen in Fig.~\ref{fig3}(d), the topmost subband is primarily localized near the center of the 5~nm QW, at a depth of $\sim2.5$~nm, which reduces its photoemission weight relative to states with stronger surface localization.

\section{Discussion}
Our results provide the first direct, momentum-resolved view of how biaxial strain and quantum confinement reshape the valence-band structure of Ge/SiGe quantum wells. By combining soft X-ray ARPES with supercell-based electronic-structure calculations, we show that the states near the valence-band maximum cannot be described as a simple heavy-hole/light-hole ladder inherited from bulk Ge. Instead, confinement and reduced symmetry generate multiple quantized subbands with momentum-dependent HH/LH/SO admixture, whose dispersion and ordering require explicit inclusion of the SiGe barrier.

This has important consequences for strained group-IV heterostructures. The familiar bulk HH, LH, and SO labels are no longer sufficient to describe the confined hole states, whose effective mass, curvature, and spin character emerge from the full heterostructure. As a result, key quantities for devices, including mobility, spin-orbit coupling, and effective $g$-factors, depend sensitively on the detailed subband composition across momentum space. Our direct determination of the Ge/SiGe valence-band offset further provides an experimental basis for engineering confinement potentials, tunnel barriers, and inter-dot couplings in Ge-based quantum devices.

More broadly, this work establishes soft X-ray ARPES, combined with realistic supercell calculations, as a powerful approach for resolving the buried electronic structure of strained semiconductor heterostructures. The ability to access momentum-resolved subband dispersions several nanometres below the surface opens a route to experimentally constrained band-structure engineering in quantum-confined materials, including Ge- and Si-based platforms and other heterostructures in which strain and confinement jointly govern the low-energy states.

\section*{Acknowledgments}
We thank Dario Theodor Marty for assistance with sample etching in the cleanroom at the Paul Scherrer Institute. We thank Nikita Shepelin and Anna Hartl for their help with XRD mapping measurements at the Paul Scherrer Institute. We acknowledge Leonard Nue for technical support during the SX-ARPES measurements at the ADRESS beamline of the Swiss Light Source (PSI). We thank Moritz Hoesch for providing the reference bulk Ge sample for band-structure measurements and for his assistance with surface preparation and measurements at the P04 beamline of PETRA III at DESY. We acknowledge Gabriel Aeppli for discussions on the topic.

\section*{Data Availability}
The data that support the findings of this study are openly available under the corresponding DOI. The simulation data presented in this work are available on the Materials Cloud platform~\cite{talirz_materials_2020} under the corresponding DOI.

\section{Methods}\label{sec4}
\renewcommand{\thefigure}{S\arabic{figure}}
\setcounter{figure}{0}

\bmhead{Heterostructure growth and structural characterization}\label{sec4.1}
The Ge/SiGe heterostructures used for this study were grown by chemical vapor deposition (CVD) through a PlasmaPro 100 Nanofab reactor equipped with a showerhead (Oxford Instruments, base pressure $<$ 0.5 mTorr), commercial germane (GeH\textsubscript{4}, Pangas, 99.999\%) and silane (SiH\textsubscript{4}, Pangas, 99.999\%) as gaseous precursors and hydrogen (H\textsubscript{2}, Pangas, 99.999\%) as diluting gas. In order to remove the native oxide, a one-minute dip in 2\% HF aqueous solution was performed on the 2 inches Si (100) substrates (float zone, undoped, resistivity $>$ 10,000 ohm$\cdot$cm) prior to growth, followed by a rinse in deionized water and isopropyl alcohol. The heterostructures were grown according to the reverse grading approach, as further detailed in \cite{Nigro2024_1,Nigro2024_2}. The growth was initiated by depositing a relaxed Ge film ($\sim$500 nm) on the Si substrate. This film, acting as a virtual substrate, was grown by following the dual-step temperature approach, involving the deposition of a thin ($\sim$100 nm) low temperature Ge seed layer (400°C, 30 mTorr GeH\textsubscript{4} partial pressure and dilution of 0.1\% GeH\textsubscript{4} in H\textsubscript{2}), followed by a thicker ($\sim$400 nm) higher temperature Ge film (500°C, 400 mTorr GeH\textsubscript{4} partial pressure and dilution of 5\% GeH\textsubscript{4} in H\textsubscript{2}). A reverse graded Si\textsubscript{1-x}Ge\textsubscript{x} alloy ($\sim$750 nm, grown at 500°C, 400 to 450 mTorr GeH\textsubscript{4} partial pressure and 5\% GeH\textsubscript{4} in H\textsubscript{2}) was subsequently deposited on the virtual substrate, by linearly decreasing the content of Ge from 100\% to 75\%. The structure was terminated by a 75\% Ge rich SiGe barrier ($\sim$300 nm, grown at 500°C, 450 mTorr GeH\textsubscript{4} partial pressure and 5\% GeH\textsubscript{4} in H\textsubscript{2}), and a thin (5 nm) Ge QW (500°C, 30 mTorr GeH\textsubscript{4} partial pressure and 1\% GeH\textsubscript{4} in H\textsubscript{2}). A 15-s long H\textsubscript{2} purging step (500°C, 100 mTorr) introduced between the growth of the two final layers, was used to enhance the sharpness at the interfaces.\\
The crystalline quality of the different materials was assessed via high-resolution Scanning Transmission Electron Microscopy (STEM) through a FEI Titan Themis.

\bmhead{Details on the ab-initio simulations}\label{sec4.2}
Kohn-Sham density functional theory is employed to perform the computational characterization of bulk Germanium.
In particular, we used the Quantum ESPRESSO package~\cite{Giannozzi2017}, which implements plane-wave basis sets and pseudopotential approach. Results are computed imposing a cutoff on the plane-wave expansion of 45 Ry and using a $12\times12\times12$ mesh for the sampling of the Brillouin zone (BZ), and spin-orbit coupling effects are included. Optimized norm-conserving Vanderbilt (ONCV) pseudopotentials \cite{Hamann2013} and the GGA-PBE exchange-correlation functional are used for all the simulations.

First, structural relaxation of the bulk and strained Ge was computed, starting from previous results in the literature (in particular as found in Terrazos \textit{et al.}~\cite{Terrazos2021}), i.e. a conventional FCC cell with lattice parameter $a = 5.658$~\AA.
Specifically, the Equation of State (EoS) was obtained by considering different values of the out-of-plane parameter c, keeping fixed the in-plane lattice parameter of the strained structure ($a=5.601$ \AA, corresponding to $\varepsilon = -1$\% strain). The results of the total energy with respect to the volume were then fitted with the Birch-Murnaghan law for the EoS. The optimal value of the out-of-plane parameter, i.e. the one minimizing the total energy curve, is found to be $c=5.884$ \AA, as shown in Fig.~S1. The resulting Poisson ratio is $\eta =-\Delta c/\Delta a=3.965$. The calculations for the EoS were performed using the AiiDA automation infrastructure~\cite{aiida2020}.

Characterization of the electronic band structure was performed using as similar setup. Additionally, hybrid functionals (HSE06)~\cite{Krukau2006} combined with spin–orbit coupling were employed to correct the unphysical gapless behavior of strained Ge within standard Kohn–Sham DFT and to reproduce the correct valence-band splittings. We used, for the primitive (conventional) cell setup, a $\textbf{k}$-point mesh sampling of $12\times12\times12$ ($8\times8\times8$) and a $\textbf{q}$-mesh of $6\times 6 \times 6$ ($4\times4\times4$) for the evaluation of the Fock exchange term, included on a fraction of 25\% as default for HSE06. 
The electronic structure was then interpolated using maximally localized Wannier functions (MLWFs)~\cite{Marzari2012,Pizzi2020}. The Wannierization was performed using 24 DFT bands and 16 target Wannier functions, considering $sp^3$ projections on the Ge atoms. 

We then used the Wannier Hamiltonian of the conventional unit cell to construct a slab Hamiltonian of 30 layers. Since the valence band structure of strained Ge and the Si$_{0.25}$Ge$_{0.75}$ substrate are very similar except for the band offset, we can mimic the heterostructure by building the slab from the bulk Hamiltonian of Ge and including an electrostatic potential in the out-of-plane direction. Concretely, we parameterize the potential by $V_\mathrm{stat}(z) = V_\mathrm{off}/2 ( 1 + \tanh[(z - z_0) / \delta] )$, where $V_\mathrm{off}$ is the band offset, while $z_0$ and $\delta$ control the position and width of the smooth transition from the top Ge layer to Si$_{0.25}$Ge$_{0.75}$. We choose $V_\mathrm{off} = -170$\, meV, $z_0 = 5$\, nm, and $\delta = 0.1 $\,nm, consistent with the experimental observations and the Poisson-Schrödinger calculations. The electrostatic potential is then incorporated into the slab Hamiltonian as described in ref.~\cite{qu_reversal_2023}:
\begin{align}
    \label{eq:slab_hamiltonian}
    H^\mathrm{slab}_{ml, m^\prime l^\prime}(\mathbf{k}_\parallel) = H^\mathrm{wan}_{m m^\prime}(\mathbf{k}_\parallel, c(l - l^\prime)) - e V_\mathrm{stat}(z_{m,l})\delta_{m m^\prime} \delta_{l l^\prime} \ ,
\end{align}
where $H^\mathrm{wan}_{m m^\prime}(\mathbf{k}_\parallel, c l)$ is the bulk Wannier Hamiltonian after partial Fourier transform over the in-plane vectors. We denote the position of each Wannier orbital in the slab by $\mathbf{r}_{m, l}$, the out-of-plane position is given by $z_{m, l} = z_m - c l$.

Diagonalizing the Hamiltonian~\eqref{eq:slab_hamiltonian} yields the layer and orbital decomposition of the Bloch states, which is used for the projections in Fig.~\ref{fig5} and for the ARPES calculations. For the latter we compute the photoemission intensity via Fermi's Golden rule,
\begin{align}
    I(\mathbf{k}_\parallel, E) = \sum_\alpha \left|M_\alpha(\mathbf{k}_\parallel, E)\right|^2 \delta(\varepsilon_\alpha(\mathbf{k}_\parallel) + \hbar\omega - E) \ ,
\end{align}
where $\varepsilon_\alpha(\mathbf{k}_\parallel) $ are the eigenvalues of Eq.~\eqref{eq:slab_hamiltonian} and where $E$ and $\hbar \omega$ denote the energy of the photoelectron and the photon energy, respectively. The photoemission matrix elements $M_\alpha(\mathbf{k}_\parallel, E)$ are expressed through the orbital decomposition as 
\begin{align}
    \label{eq:matrix_elements}
    M_\alpha(\mathbf{k}_\parallel, E) = \sum_{m, l} U_{ml, \alpha}(\mathbf{k}_\parallel) e^{-i\mathbf{p}\cdot\mathbf{r}_{m,l}} e^{z_{m,l} / \lambda} M_{m}(\mathbf{k}_\parallel, E) \ .
\end{align}
Here, $\mathbf{p}$ denotes the photoelectron momentum ($\mathbf{p}^2 / 2m = E$) with $\mathbf{p}_\parallel = \mathbf{k}_\parallel$, while the out-of-plane momentum is determined from $E$ and the inner potential $V_{000}$. For simplicity we replace the matrix in orbital basis $M_{m}(\mathbf{k}_\parallel, E) \rightarrow 1$. This approximation removes the polarization dependence, but retains all structural interference effects through Eq.~\eqref{eq:matrix_elements}. The surface sensitivity is controlled by the mean-free path $\lambda$, which we fix at 2 nm.

The spectral functions in Fig.~\ref{fig5} were computed by projecting the slab bands onto the bulk states. To this end we calculated the eigenvectors of the bulk Wannier Hamiltonian, $U^\mathrm{bulk}_{m,n}(\mathbf{k}_\parallel, k_z)$ and stack them according to $\tilde{U}^\mathrm{bulk}_{ml,n}(\mathbf{k}_\parallel, k_z) = U^\mathrm{bulk}_{m,n}(\mathbf{k}_\parallel, k_z) e^{i k_z c l}$. In the limit of very large slab size, approaching the periodic limit, $\tilde{U}^\mathrm{bulk}_{ml,n}(\mathbf{k}_\parallel, k_z)$ becomes independent of $k_z$ and represents the eigenvectors of the slab Hamiltonian. Thus, by projecting the finite-size slab eigenvectors $U_{ml,\alpha}(\mathbf{k}_\parallel)$ onto $\tilde{U}^\mathrm{bulk}_{ml,n}(\mathbf{k}_\parallel, k_z)$, we can extract the content of the bulk bands in the slab electronic structure. Concretely we define the spectral weights
\begin{align}
    \label{eq:spectral_weights}
    S_{n\alpha}(\mathbf{k}_\parallel) = \sum_{l,m} w_l \left| [\tilde{U}^\mathrm{bulk}_{ml,n}(\mathbf{k}_\parallel, k_z)]^* U_{ml,\alpha}(\mathbf{k}_\parallel) \right|^2 \ ,
\end{align}
where the weights $w_l = 1$ for the top 5 nm ($w_l = 0$ otherwise). Thus, the spectral weights~\eqref{eq:spectral_weights} also capture the projection onto the Ge QW. Finally, we compute the spectral function
\begin{align}
    \label{eq:spectral_function}
    A_{n}(\mathbf{k}_\parallel, \omega) = \sum_{\alpha} S_{n,\alpha} g(\omega - \varepsilon_\alpha(\mathbf{k}_\parallel)) \ ,
\end{align}
where $g(\omega)$ denotes a Gaussian $g(\omega) = e^{-\omega^2/2\sigma^2} / \sqrt{2\pi \sigma^2}$ with $\sigma=50$~meV. The spectral function~\eqref{eq:spectral_function}, resolved with respect to bulk bands $n$, is plotted in Fig.~\ref{fig5}.

\begin{figure}[H]
\centering
\includegraphics[width=0.8\textwidth]{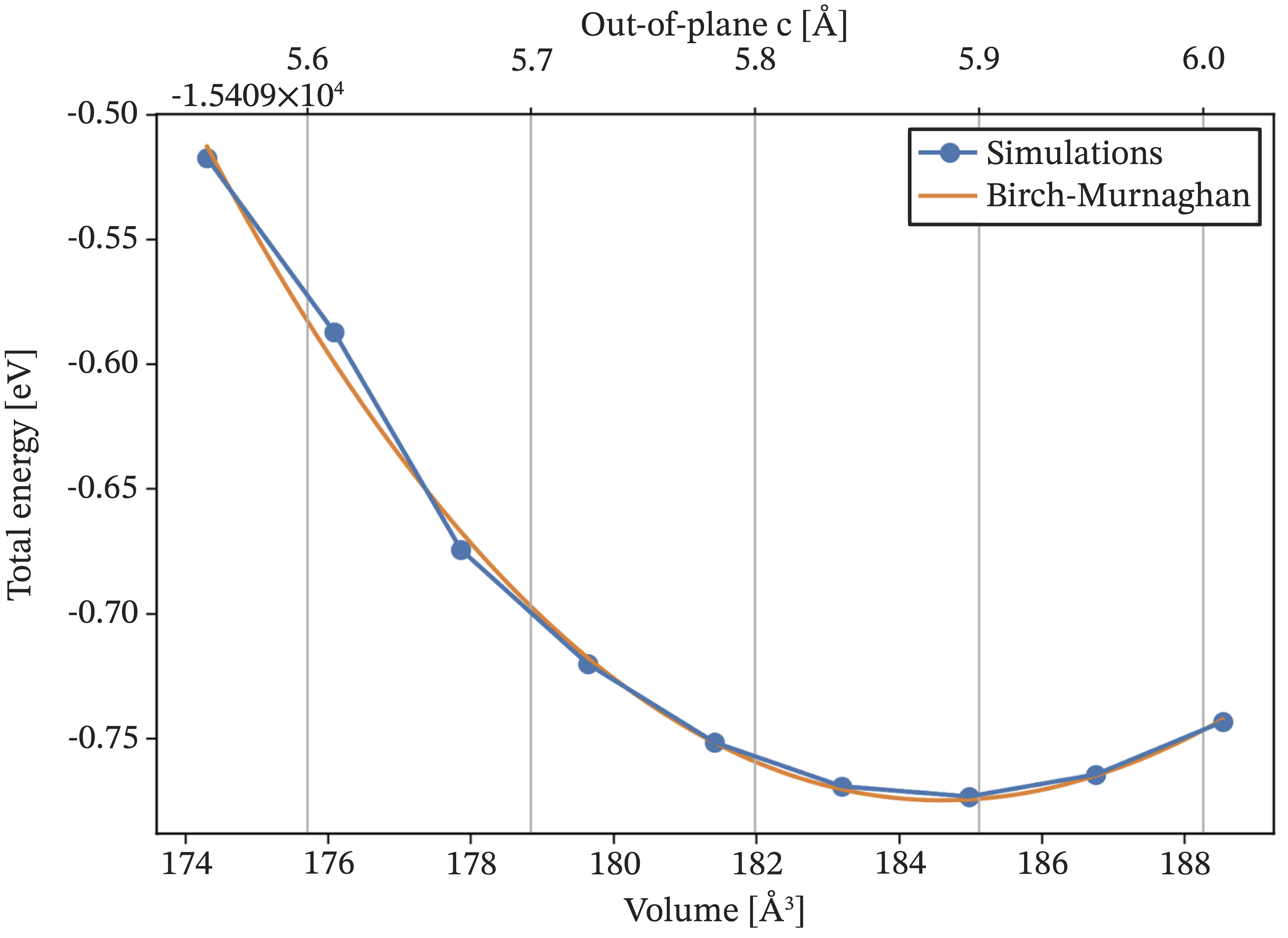}
\caption{\textbf{Equation of state for bulk Ge.} Total energy versus the volume of the strained structure, related to its out-of-plane lattice vector $c$.
}
\label{figS1}
\end{figure}

\begin{figure}[H]
\centering
\includegraphics[width=1\textwidth]{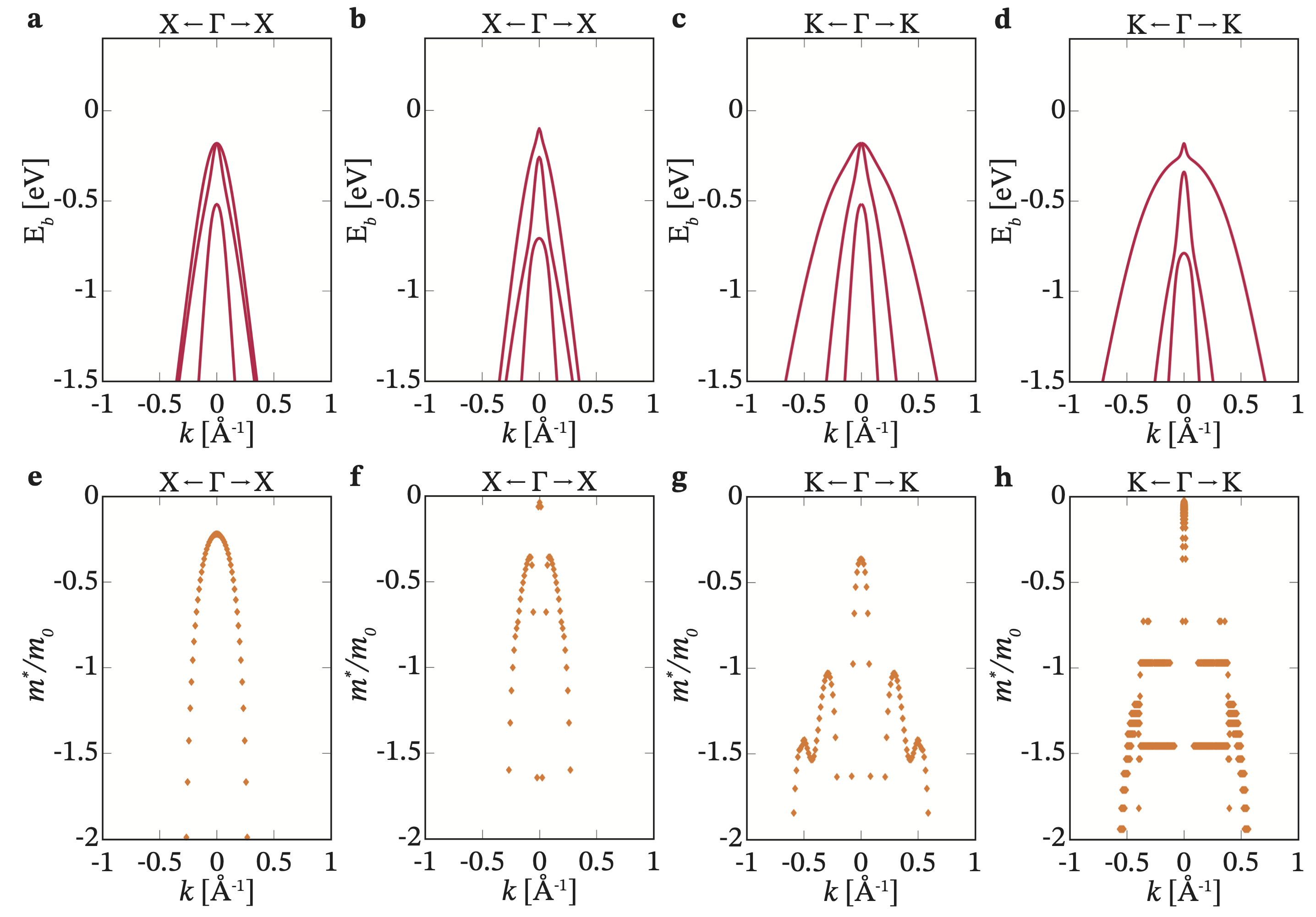}
\caption{\textbf{Bulk electronic structure and effective mass of (un)strained Ge.} (a), (b) and (f), (g) calculated bulk band structures of bulk relaxed and $\varepsilon = -1\%$ strained germanium, respectively, using the Heyd-Scuseria-Ernzerhof (HSE) hybrid density functionals. (c), (d) and (h), (i) computed effective mass of the topmost valence band derived from the HSE band structures for unstrained and strained conditions. (e) Illustration of the first Brillouin zone highlighting key high-symmetry points and directions.}
\label{figS2}
\end{figure}

\bmhead{Surface preparation for ARPES measurements}\label{sec4.3}
The surface preparation of all crystals measured by SX-ARPES in this work followed a standardized procedure. Before measurement, the crystals were dipped in hydrofluoric acid (HF) with a concentration of 1\% for 60~s to remove the oxide layer formed upon air exposure. Subsequently, the crystals were rinsed in high-purity isopropanol. The crystals remained in isopropanol for approximately 5~min and were then transferred directly into ultrahigh vacuum through a load lock, during which the surface remained covered by isopropanol until the pressure reached approximately $10^{-4}$~mbar.

\bmhead{SX-ARPES measurements}\label{sec4.4}
Most SX-ARPES measurements were performed at the SX-ARPES endstation~\cite{Strocov2014} of the ADRESS beamline~\cite{Strocov2010} at the Swiss Light Source (Paul Scherrer Institut), while additional measurements were carried out at the same endstation at the P04 beamline of PETRA III (DESY). The available photon energies spanned from $h\nu = 300$~eV to $h\nu = 1600$~eV (and from $h\nu = 250$~eV to $h\nu = 3000$~eV at P04), with a photon flux on the order of $10^{13}$~ph/s/(0.01\% BW). The experiments were conducted at a cryogenic temperature of 10~K to minimize thermal disorder effects on the ARPES spectra. The in-plane photoelectron momentum $k_x$ is determined from the emission angle along the analyzer slit, while the photoelectron momentum $k_y$ is determined from the emission angle perpendicular to the analyzer slit using electric-field deflection; as mentioned in the main text, $k_z$ is obtained by varying $h\nu$. The angular resolution of the PHOIBOS-225 analyzer is better than $0.1^\circ$, while the combined energy resolution of the beamline and analyzer under typical measurement conditions at $h\nu \sim 460$~eV is 45~meV. Further details of the SX-ARPES endstation, including the experimental geometry, can be found in Ref.~\cite{Strocov2014}.

\backmatter






\bibliography{sn-bibliography}

\end{document}